\documentclass{aastex631}
\usepackage{graphicx}
\usepackage{color}

\shorttitle{}
\shortauthors{}

\begin{document}

\title{Quasar standardization: Overcoming Selection Biases and Redshift Evolution}

\author{Maria Giovanna Dainotti}
\affiliation{National Astronomical Observatory of Japan, 2 Chome-21-1 Osawa, Mitaka, Tokyo 181-8588, Japan}
\affiliation{The Graduate University for Advanced Studies, SOKENDAI, Shonankokusaimura, Hayama, Miura District, Kanagawa 240-0193, Japan}
\affiliation{Space Science Institute, Boulder, CO, USA}
\author{Giada Bargiacchi}
\affiliation{Scuola Superiore Meridionale,
                Largo S. Marcellino 10,
                80138, Napoli, Italy}
\affiliation{Istituto Nazionale di Fisica Nucleare (INFN), Sez. di Napoli,
                Complesso Univ. Monte S. Angelo, Via Cinthia 9,
                80126, Napoli, Italy}

\author{Aleksander Łukasz Lenart}
\affiliation{Astronomical Observatory, Jagiellonian University, ul. Orla 171, 31-501 Kraków, Poland}

\author{Salvatore Capozziello}
\affiliation{Dipartimento di Fisica "E. Pancini" , Universit\'a degli Studi di  Napoli "Federico II"\\
                Complesso Univ. Monte S. Angelo, Via Cinthia 9
                80126, Napoli, Italy}
\affiliation{Scuola Superiore Meridionale, 
                Largo S. Marcellino 10,
                80138, Napoli, Italy}
\affiliation{Istituto Nazionale di Fisica Nucleare (INFN), Sez. di Napoli,
                Complesso Univ. Monte S. Angelo, Via Cinthia 9,
                80126, Napoli, Italy}
\author{Eoin \'O Colg\'ain}
\affiliation{Center for Quantum Spacetime, Sogang University, Seoul 121-742, Korea}
\affiliation{Department of Physics, Sogang University, Seoul 121-742, Korea}

\author{Rance Solomon}
\author{Dejan Stojkovic}
\affiliation{HEPCOS, Department  of  Physics,  SUNY  at  Buffalo,  Buffalo,  NY  14260-1500, USA}
\author{M.M. Sheikh-Jabbari}
\affiliation{School of Physics, Institute for Research in Fundamental Sciences (IPM), P.O.Box 19395-5531, Tehran, Iran}


\begin{abstract}
Quasars (QSOs) are extremely luminous active galatic nuclei currently observed up to redshift $z=7.642$. As such, they have the potential to be the next rung of the cosmic distance ladder beyond SNe Ia, if they can reliably be used as cosmological probes. The main issue in adopting QSOs as standard candles (similarly to Gamma-Ray Bursts) is the large intrinsic scatter in the relations between their observed properties. This could be overcome by finding correlations among their observables that are intrinsic to the physics of QSOs and not artifacts of selection biases and/or redshift evolution. The reliability of these correlations should be verified through well-established statistical tests. The correlation between the ultraviolet (UV) and X-ray fluxes developed by Risaliti \& Lusso is one of the most promising relations. We apply a statistical method to correct this relation for redshift evolution and selection biases. Remarkably, we recover the the same parameters of the slope and the normalization as Risaliti \& Lusso. Our results establish the reliability of this relation, which is intrinsic to the QSO properties and  not merely an effect of selection biases or redshift evolution. Hence, the possibility to standardize QSOs as cosmological candles, thereby extending the Hubble diagram up to $z=7.54$.

\end{abstract}


\section{Introduction} \label{sec:intro}
The quest for standard candles at high redshifts is still open with the aim of extending the Hubble diagram out beyond the epoch of reionization. Since the discovery of Gamma-Ray Bursts (GRBs) as extragalactic sources, 
hosts of high redshift sources have been identified, 
including GRB 090423 \citep{2009arXiv0906.1577T}  and  GRB 090429B \citep{2011ApJ...736....7C}.
Recently,  quasars (QSOs) have also been observed at high redshifts, reaching up to $z=7.54$ \citep{banados2018} and $z=7.64$ \citep{2021ApJ...907L...1W}. One of the biggest challenges for the use of these objects as standardizable candles is the large scatter in the relations among their intrinsic properties. Since 2002, many authors in the GRB community have investigated the possibility of using GRB relations as cosmological probes \citep{Dainotti2008,2010ApJ...722L.215D, dainotti11a,Dainotti11b,Dainotti2013a,Dainotti2013b,Dainotti2015b,2015ApJ...800...31D,2016ApJ...825L..20D,2017A&A...600A..98D,2017NewAR..77...23D,2018AdAst2018E...1D,2018PASP..130e1001D, 2020ApJ...904...97D,2020ApJ...905L..26D,Srinivasaragavan2020,2021Galax...9...95D,2021PASJ...73..970D,2021ApJ...914L..40D,2021ApJS..255...13D,2022arXiv220105245C,2022MNRAS.510.2928C,cardone10,postnikov14}.
The search for high redshift standard candles has been boosted by the Hubble tension, a $4-6$ $\sigma$ discrepancy between the direct measurements of local $H_0$ and the one inferred from cosmological models, most notably the value reported by the Planck observation within $\Lambda$CDM model. Additional high-$z$ standardized probes beyond SNe Ia, such as GRBs and QSOs, could be instrumental in shedding light on this problem \citep{Spallicci,Rocco,2021A&A...649A..65B,2021ApJ...912..150D,2022arXiv220109848D, Moresco:2022phi}.


QSOs are extremely luminous active galactic nuclei (AGNs). Their emission cannot be explained by standard stellar processes and requires a different kind of mechanism, e.g. mass accretion onto the central supermassive black hole \citep[see e.g.][]{1998A&A...334...39S,1999RvMPS..71..180H,qsophysics,2020MNRAS.498.5652K}. This mechanism can indeed explain the observed properties of QSO emission, especially (for what concerns our interest) the UV and X-ray emissions. The accretion disk emits photons in the UV band, which are then processed through inverse Compton effect by an external plasma of relativistic electrons, giving rise to X-ray emission. This physical explanation, while plausible, 
lacks in accounting for the stability of the X-ray emission. Ultimately, one needs an efficient energy transfer between the accretion disk and the external relativistic `corona' to explain such a stable emission. The physical origin of this link between the two AGN regions is not known yet. However, some models have been proposed \citep[see e.g.][]{2017A&A...602A..79L} yielding relations that have been confirmed by the empirical correlation between UV and X-ray QSO luminosities.
One of the most remarkable QSO correlations proposed so far is the so-called Risaliti-Lusso relationship among the fluxes in UV and X-ray bands, based on the non-linear relation between their UV and X-ray luminosities \citep{1979ApJ...234L...9T,1982ApJ...262L..17A,1985ApJ...297..177K,1986ApJ...305...83A,2003AJ....125..433V,steffen06,2007ApJ...665.1004J, 2010A&A...512A..34L,lr16,2021arXiv210903252B}. The relation is extremely powerful because, if true, it allows one to standardize QSOs across a wide range of luminosities. 
This relation has been applied as a cosmological tool and more in general the QSOs community is currently investigating the application of QSOs with other methods as cosmological tools and the possible problems associated with them  \citep[e.g.][]{2021arXiv211200052K,2021MNRAS.508.4722K}.
In terms of luminosities, the Risaliti-Lusso relation may be expressed as\footnote{For the sake of simplicity we always use $\text{log}$ instead of $\text{log}_{10}$.} 
\begin{equation}\label{RL-LL}
\mathrm{log} L_{X} = \gamma \, \mathrm{log} L_{UV} + \beta
\end{equation}
where  $\beta,\gamma$ are (constant) fitting parameters and 
\begin{equation}\label{L-F--Dl}
L_{X,UV}= 4 \pi D_{L}^{2}\ F_{X, UV}  \label{luminosities} 
\end{equation}
where UV and X refer to 2500 \AA \, and 2 KeV, respectively. 

We here point out that the formula is written in a way for which the $L_X$ is derived by $L_{UV}$ since are the UV photons emitted by the accretion disc that represent the "seeds" for the X-ray emission of the corona through inverse-Compton scattering. Indeed, if one can turn-off the disc the corona will immediately follow, but the opposite is not true. If one can turn-off the corona, the disc will still emit its luminosity regardless.

We note that luminosities are obtained applying a K-correction.
The K-correction is defined as $1/(1+z)^{1-\alpha}$ where $\alpha$ is the spectral index of the sources and it is  assumed to be 1 for the sources leading to a $K=1$, so hereafter the K-correction has been omitted following \citet{2020A&A...642A.150L}.
In Eq.(\ref{RL-LL}), once we substitute Eq.(\ref{luminosities}), the dependence through the luminosity distance $D_{L}$ becomes evident.  This relation has been confirmed using various samples of QSOs, but with a very large intrinsic dispersion, hereafter denoted with $\delta$, $\delta\sim 0.35/0.40$ dex in logarithmic units \citep[e.g.][]{2010A&A...512A..34L}. Only recently, it has been pointed out that this dispersion has mainly an observational and non-intrinsic origin \citep{lr16}. This finding has allowed to reduce the intrinsic scatter to $\delta\sim 0.2$ dex and has rendered this relation suitable for cosmological analyses, turning QSOs into reliable cosmological tools. We refer to \citet{lr16}, \citet{rl19} and \citet{2020A&A...642A.150L} for a more detailed description on the physics of this relation and on its cosmological use. 

This method has been developed only very recently and still needs to be tested and checked against different possible issues mainly related to selection biases, dependence of the relation on the black hole mass and accretion rate in the AGN, and redshift evolution; thus further tests are needed to probe its reliability. Some of its issues are highlighted in
\citet{Yang:2019vgk} and \citet{2021MNRAS.502.6140K,2022MNRAS.510.2753K}. 

To fully cast light on the intrinsic nature of this relation, here we perform for the first time in the literature its correction for selection biases and redshift evolutionary effects with reliable statistical methods.
If such biases were present, they could invalidate its reliability from a physical point of view and as a cosmological application. 
More precisely, if the correlation was merely induced by selection biases and redshift evolution, the slope of the correlation after correction for the biases and the evolution should have been  compatible with a slope=0 within 5 $\sigma$. This is not the case, and as we demonstrate here,  the slope of the Risaliti-Lusso relation is not compatible with zero even at the 52.9 $\sigma$ level after we apply the corrections. 
In this paper, we explore the standardization of QSOs in view of future cosmological applications. In Sec. \ref{sample}, we discuss in detail the QSO sample we use. Sec. \ref{method} is devoted to the statistical analysis and the discussion of both selection biases and redshift evolution. In Sec. \ref{correlation}, we consider the Risaliti-Lusso correlation with the aim of verifying its intrinsic nature. In Sec. \ref{conclusion} we summarize our results and discuss future perspectives.

\section{The sample}
\label{sample}

We use the most up-to-date sample of Risaliti-Lusso QSOs \citep{2020A&A...642A.150L}. This is composed of 2421 sources in the redshift range $z =0.009-7.54$. These sources have been carefully selected for cosmological studies addressing possible observational issues, such as dust reddening, host-galaxy contamination, X-ray absorption and Eddington bias, as detailed in \citet{2020A&A...642A.150L}. In particular, all QSOs with a spectral energy distribution (SED) that show reddening in the UV and significant host-galaxy contamination in the near-infrared, are removed, leaving only sources with extinction E(B$-$V)$\leq 0.1$. 
This requirement is fulfilled by selecting only the sources that satisfy $\sqrt{(\Gamma_{1,\mathrm{UV}}-0.82)^2 + (\Gamma_{2,\mathrm{UV}}-0.40)^2} \leq 1.1$, where $\Gamma_{1, \mathrm{UV}}$ and $\Gamma_{2, \mathrm{UV}}$ are the slopes of a log($\nu$)-log($\nu \, L_\nu$) power-law in the rest frame 0.3-1$\mu$m and 1450-3000 \AA \, ranges respectively, and $\nu$ and $L_\nu$ denote the frequency and the luminosity per unit of frequency. The specific values  $\Gamma_{1, UV}=0.82$ and $\Gamma_{2, UV}=0.4$ refer to a SED with zero extinction. In addition, X-ray observations where photon indices ($\Gamma_X$) are peculiar or indicative of X-ray absorption are excluded by requiring $\Gamma_X + \Delta \Gamma_X \geq 1.7$ and $\Gamma_X \leq 2.8 $ if $z < 4$ and $\Gamma_X \geq 1.7$ if $z \geq 4$, where $\Delta \Gamma_X$ is the uncertainty on the photon index. Finally, the remaining observations are filtered to correct for the Eddington bias. 

The final cleaned sample is hence composed only of sources satisfying $\text{log}F_{X,exp} - \text{log}F_{min} \geq {\cal F}$, where ${\cal F}$ stands for a filtering threshold value and
$F_{X,exp}$ is the X-ray flux  expected from the observed UV-flux assuming the Risaliti-Lusso $F_X - F_{UV}$ relation with fixed $\gamma$ and $\beta$ within the flat $\Lambda${CDM} model with $\Omega_{M}=0.3$ and $H_{0} = 70\, \mathrm{km\,s^{-1}\,Mpc^{-1}}$. $F_{lim}$ is the flux limit of the specific observation estimated from the catalogue. 
The value of ${\cal F}$ required in this filter is $0.9$ for the Sloan Digital Sky Survey SDSS–4XMM and XXL subsamples and  $ 0.5$ for the SDSS-\textit{Chandra}. 

For any QSO, all the multiple X-ray observations that survive the filters above are finally averaged to minimize the effects of X-ray variability. The cleaned sample used in this work is the product of all these selection criteria.




\section{Statistical analysis to overcome selection biases and redshift evolution}
\label{method}
We apply the 
\citet{1992ApJ...399..345E} (EP) statistical method, which is able to correct for selection biases and redshift evolution, thereby uncovering intrinsic correlations in extragalactic objects (such as QSOs and GRBs). The reliability of this procedure has already been demonstrated via Monte Carlo simulations for GRBs \citep{Dainotti2013b}. We detail the method in Appendix \ref{appendix}, while here we only summarize the crucial points required by the present work.

Following the approach in \citet{Dainotti2013a,Dainotti2015b,2017A&A...600A..98D,2021Galax...9...95D}, we can correct for the evolution and obtain the local variables, in our case the luminosities. These new variables denoted with $'$ are the so-called ``de-evolved variables", since the evolution has been removed. Such a removal may be achieved using the function $g(z)=(1+z)^{k}$ where the $k$ parameter mimics the evolution with redshift and the new, de-evolved luminosities $L'$ are obtained from the original $L$ via $L' = L/g(z)$. 
The functional form for $g(z)$ can be a simple power-law \citep{Dainotti2013a,2017A&A...600A..98D}, similarly to this one, or a more complex function such as 
$g(z)=(Z^k \, z^k_{crit})/(Z^k+z^k_{crit})$ shown in \cite{2011ApJ...743..104S}, where $Z=1+z$, 
which allows for a more rapid evolution up to redshift $z_{crit}$ than the less rapid one at higher redshifts. This is a good fit for a QSO dataset based on SDSS with many QSOs 
at $z > 3$ \citep{2013ApJ...764...43S,2016ApJ...831...60S,2019ApJ...877...63S}. 
In this work, we use the latter form with a fiducial critical redshift $z_{crit}=3.7$, which is the most suitable value given the high redshift distribution of the QSOs determined in \citet{2013ApJ...764...43S},
but we also check our results against the specific functional form allowing also for a simple power-law. Interestingly, the same $g(z)$ was shown to  reproduce the observed luminosity function of AGNs \citet{2013ApJ...764...43S}.

Here, we detail our results of the EP method for the studied parameters for the whole sample of 2421 QSOs considering the evolutionary form with $z_{crit}=3.7$. 
The EP method takes into account both possible biases from incomplete data and redshift evolution of observables by using an adaptation of the Kendall $\tau$ statistic (see \citealt{Dainotti2013b} and Sec. 7 of \citealt{Dainotti2015b} for a more detailed description). This test allows determining the correlation between two generic variables $(x_i,y_i)$ and the best-fit values of parameters describing their correlation function by defining the parameter $\tau$ as
\begin{equation}\label{tau}
\tau =\frac{\sum_{i}{(\mathcal{R}_i-\mathcal{E}_i)}}{\sqrt{\sum_i{\mathcal{V}_i}}}
\end{equation}
where $\mathcal{R}_i$ is the rank of $y_i$ in a set associated with it, $\mathcal{E}_i$ and $\mathcal{V}_i$ are its expectation value and variance, respectively. For untruncated data the associated set includes all of the data with $x_j < x_i$. 
In our case in which we have truncation, the associated set for $z_i$ contains all QSOs with $L_{z_j} \geq  L_{min,i}$ and $z_j \leq z_i$. Here $j$ and $i$ denote objects of the associated set and the overall QSO sample, respectively, and $L_{min}(z_i)$ is the minimum luminosity that would still allow us to detect an object at a given $z_i$.
More specifically, $\mathcal{R}_i$ is computed for each data point considering the position of the data in samples including all the objects that can be detected considering particular observational limits, see \cite{1992ApJ...399..345E} for further details.  If two variables are independent, $\mathcal{R}_i$ should be distributed continuously between $0$ and $1$ with $\mathcal{E}_i=(i+1)/2$ and $\mathcal{V}_i=(i^{2}-1)/12$. Independence is rejected at $n \sigma$ level if $| \tau | > n$. With this statistic, we find the parametrization that best describes the evolution. For the present study, we are interested in analyzing the redshift evolution of QSO UV and X-ray luminosities to test their degree of correlation in the $(z,\mathrm{log}L'_{UV})$ and $(z,\mathrm{log}L'_{X})$ spaces. To eliminate the depence of the redshift on our variables of interest, we require $\tau(k)=0$, which implies 
the de-evolved luminosities are statistically independent of the redshift.
The value of $k$ corresponding to $\tau=0$ gives us the exact redshift evolution of $L_{UV}$ and $L_{X}$ within 1 $\sigma$, determined from $|\tau| \leq 1$, according to our chosen functional form for $g(z)$. Here, we detail the computations to derive the evolutionary coefficients and follow the same procedure for both $L_{UV}$ and $L_{X}$.

From the measured flux we compute the luminosity for each QSO assuming a flat $\Lambda \mathrm{CDM}$ model with $\Omega_{M}=0.3$ at the current time and $H_{0} = 70  \, \mathrm{km\,s^{-1}\,Mpc^{-1}}$. Note that in another investigation (Dainotti et al. 2022, MNRAS submitted) it is shown that the values of the evolutionary parameters for GRBs do not evolve with the cosmological parameters such as $\Omega_{M}$ and $H_0$. Thus, we can safely use this test with any given cosmological model and we explicitly use this test later in Sec. \ref{impact on cosmology of the g(z)} . 
There is also an ongoing discussion on the validity of the Risaliti-Lusso relation and its effectiveness beyond $z \sim 1.5$ due to discrepancies between QSQs and SNe Ia; for example, see \citet{Yang:2019vgk} and \citet{2021MNRAS.502.6140K,2022MNRAS.510.2753K}. 
However, this discussion is beyond the scope of the current paper.

We also compute the flux limit $F_{lim}$ and the corresponding luminosity $L_{min}(z_i)$. According to \citet{Dainotti2013a}, \citet{Dainotti2015b}, \citet{2017A&A...600A..98D}, \citet{2022ApJ...925...15L} and \citet{2022arXiv220315538D}, the samples used to derive the evolutionary effects should not be less than the $90 \%$ of the original ones and the population of X-rays and UV should resemble as much as possible to the overall distribution. To this end, conservative choices regarding the limiting values are needed. 

Specifically, we have chosen $F_{lim} = 4.5\times \, 10^{-29} \,\mathrm{erg \, s^{-1} \, cm^{-2} \, Hz^{-1}}$ for the UV and $F_{lim} = 6\times \, 10^{-33}\, \mathrm{erg \, s^{-1} \, cm^{-2} \, Hz^{-1}}$ for the X-rays, which respectively guarantee samples of 2362 (97.6\%) and 2379 (98.3\%) QSOs. 
We have also verified through the means of the Kolmogorov Smirnov (KS) Test that the full and the cut samples in both X-rays and UV come from the same parent population. Indeed, the probability of the null-hypothesis that the two samples are drawn by the same distribution cannot be rejected at the $p=0.47$ for the UV and $p=0.85$ for X-rays. 

\begin{figure*}[h!]
\gridline{\fig{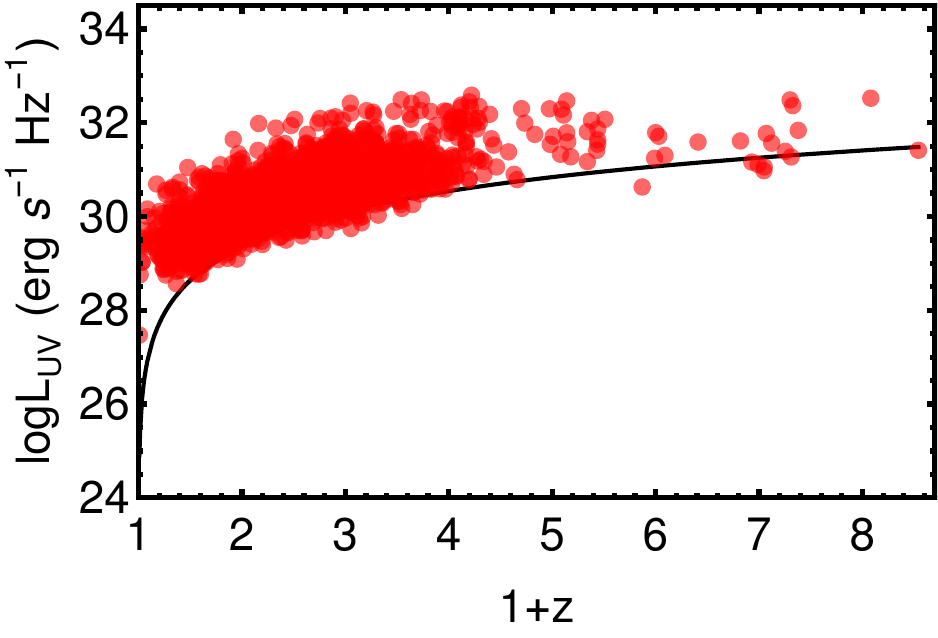}{.5\textwidth}{}
            \fig{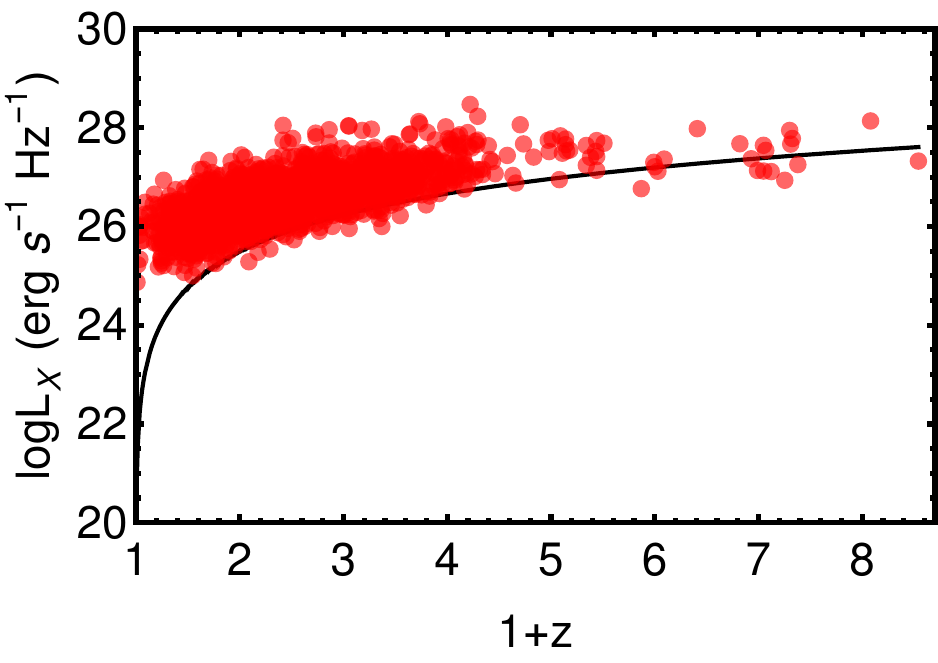}{.5\textwidth}{} }
    \caption{Redshift evolution of $L_{UV}$ (left panel) and $L_{X}$ (right panel) in units of $\mathrm{erg \, s^{-1}\, Hz^{-1}}$ for the whole QSO sample (red circles). The black line in both panels shows the limiting luminosity chosen according to the prescription described in Sect.\ref{method}.}
    \label{fig:evoL-z}
\end{figure*}

\begin{figure*}[h!]
\gridline{\fig{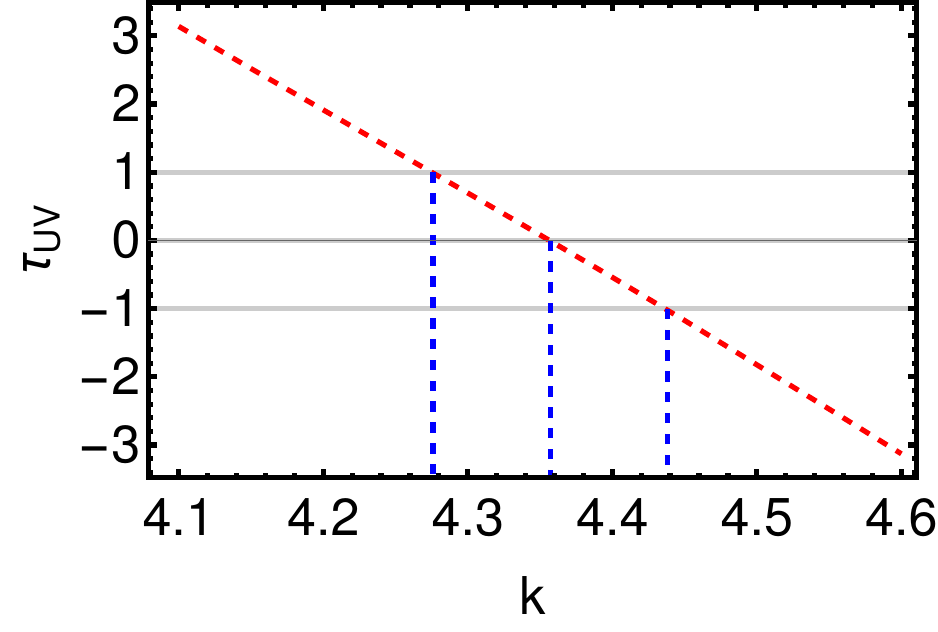}{.5\textwidth}{}
            \fig{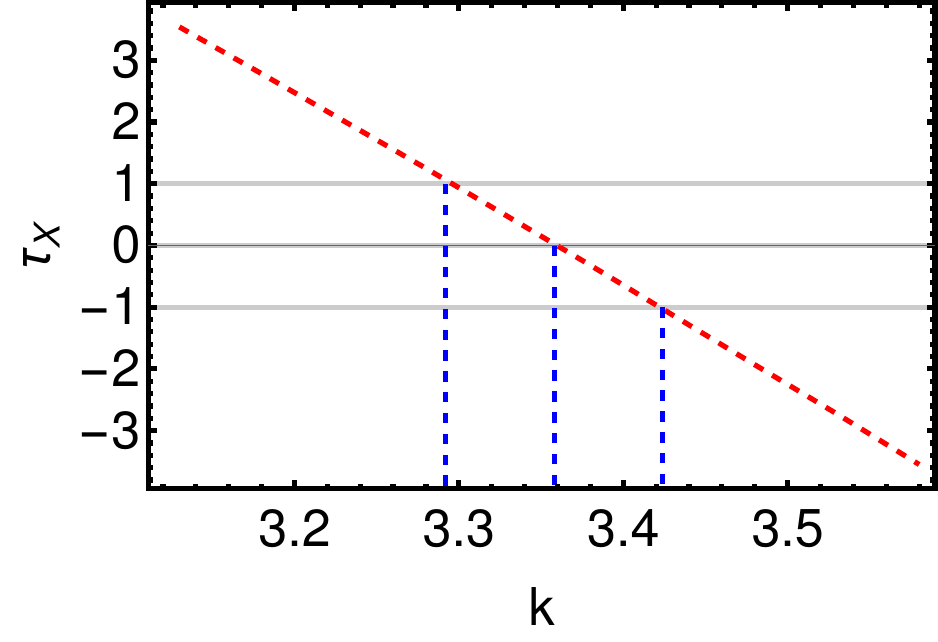}{.5\textwidth}{} }       
    \caption{$\tau (k)$ function (dashed red line) for both the UV (left panel) and X-ray (right panel) analyses. The point $\tau = 0$ gives us the $k$ parameter for the redshift evolution of $L_{UV}$ and $L_{X}$, while $|\tau| \leq 1$ (gray lines) the 1 $\sigma$ uncertainty on it (dashed blue lines). }
    \label{fig:evot-k}
\end{figure*}

The limiting values for $L_{UV}$ and $L_X$ corresponding to these values of $F_{lim}$ are shown with a black continuous line in the left and right panel of Fig.\ref{fig:evoL-z}, respectively, over the whole set of data points represented by red circles. 
We then apply the $\tau$ test to the data sets trimmed with values of the fluxes mentioned above and obtain the trend for $\tau(k)$ shown in the left and right panel of Fig. \ref{fig:evot-k} for the UV and X-rays, respectively. As already explained, $\tau = 0$ and $|\tau| \leq 1$ provide us with the best-fit value and the associated 1 $\sigma$ error for the evolutionary parameter $k$. For the UV and X-rays we obtain $k= 4.36 \pm 0.08$ and $k=3.36 \pm 0.07$, respectively. It is remarkable that the evolutionary function of the UV in our sample is compatible within $2.4$ $\sigma$ with the optical evolutionary coefficient $k_{opt}$ obtained in \citet{2013ApJ...764...43S}, where the same form of $g(z)$ is used. In their paper they found $k_{opt}=3.0 \pm 0.5$ and corrected the luminosity function with the central value. Thus, the new luminosity function can be representative of the observed luminosity function, but it will be constructed with the local luminosities (de-evolved luminosities), and thus, they will be rescaled by the $g(z)$ functions. Indeed, similarly to \citet{2013ApJ...764...43S} we expect that the results of our luminosity function are in agreement with the ones in the literature.

However, we note that if a different method regarding the choice of the limiting luminosity is applied the evolutionary functions are smaller \citep[see][]{2022arXiv220313374S}. 
The method detailed in \citep[see][]{2022arXiv220313374S} takes into consideration a different limit for each source as $F_{j,x,lim}=F_{j,x} \times \sigma_{j,min}/\sigma_j$ where the ratio of an object $j$'s indicates the significance $\sigma_j$ to its minimum significance $\sigma_{j,min}$. This ratio is used to calculate the minimum X-ray flux for each source. 
In addition, we may note that in \citet{2022arXiv220313374S} the K-correction has been applied to each source, while in our case the K-correction is assumed to be 1. Another major difference is that the sample used in the case of \citet{2022arXiv220313374S} is taken from the SDSS DR7 \citep{2010AJ....139.2360S}, instead in our case we use a sample of 2421 sources from the newest release SDSS DR14 \citep{2018A&A...613A..51P}. 
It is definitely interesting to consider this more complex approach with the flux limits in a forthcoming paper.

We have performed an additional test to legitimate our choice for $F_{lim}$ and prove that the evolutionary coefficients depend only weakly on these choices. In both X-ray and UV band, we compute the evolutionary coefficient k for different limiting values $F_{lim}$.
Specifically, we started with a value of $F_{lim}$ that preserves the original sample and then analyzed a range of values for $F_{lim}$. If we span within 0.5 magnitude in the UV starting from $F_{lim, UV} = 1.7\times \, 10^{-29} \,\mathrm{erg \, s^{-1} \, cm^{-2} \, Hz^{-1}}$ and in X-rays starting from $F_{lim, X-rays} = 1.5\times \, 10^{-33} \,\mathrm{erg \, s^{-1} \, cm^{-2} \, Hz^{-1}}$ we obtain a compatibility within 1 $\sigma$. Even if we span over one order of magnitude starting from the same $F_{lim}$ values in fluxes both in UV and in X-rays, the evolutionary coefficient results remain compatible within 2 $\sigma$. 
%
This analysis proves that the results for the evolutionary coefficients do not depend on the specific choice of $F_{lim}$ for a wide range of their values.

\begin{figure*}[h!]
\plotone{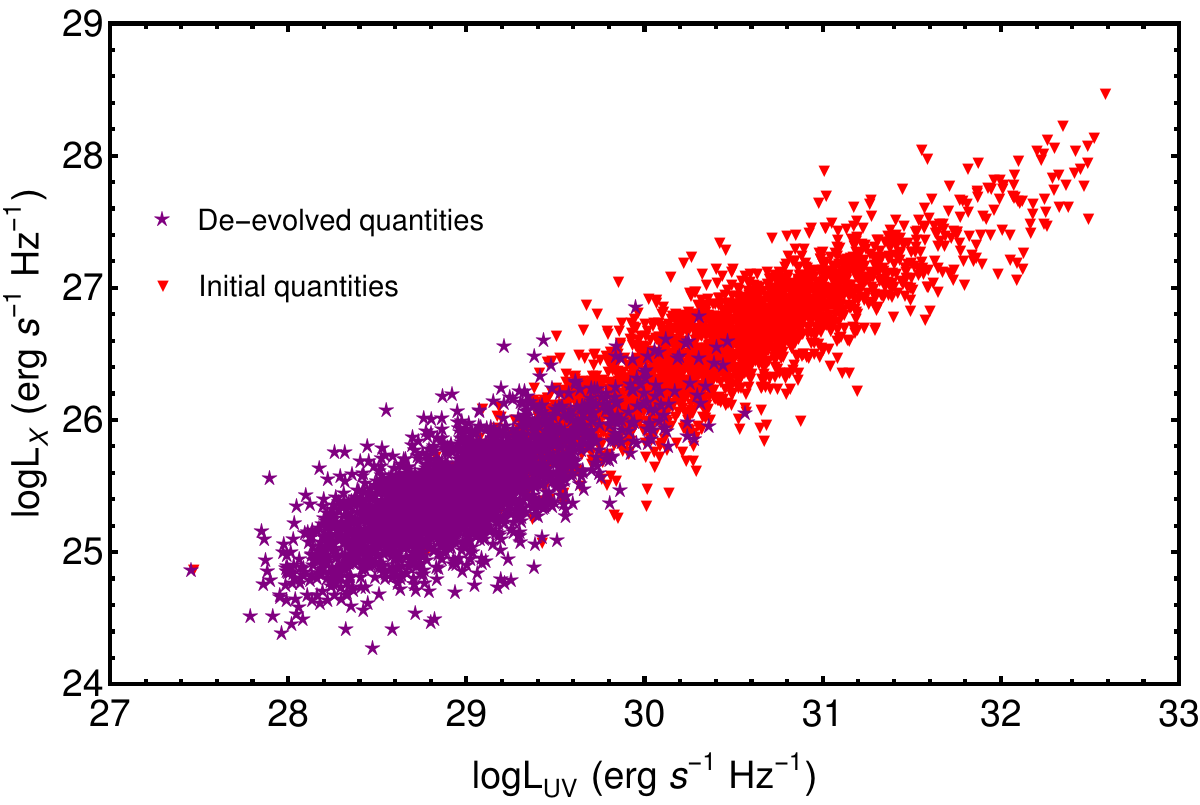}
    \caption{Comparison between initial (red) and de-evolved (purple) quantities in the ($\mathrm{log}L_{UV}$, $\mathrm{log}L_{X}$) plane.}
    \label{deev}
\end{figure*}

Inserting our values of $k$ in $g(z)$, we then compute the new de-evolved luminosities, denoted with $'$, and the associated uncertainties for the whole original QSO sample. The comparison between these quantities and the initial ones is shown in Fig.\ref{deev} in the ($\mathrm{log}L_{UV}$, $\mathrm{log}L_{X}$) plane. Compared to the initial ones, the computed luminosities span a smaller region of the ($\mathrm{log}L_{UV}$, $\mathrm{log}L_{X}$) plane and show a slightly greater dispersion ($\delta \sim 0.22$ against $\delta \sim 0.21$, as evaluated in Sec.\ref{correlation}). This fact is expected because the $g(z)$ function, once the best-fit values for $k$ are used, yields a greater correction (i.e. lower de-evolved values) for higher luminosities. In addition, we have accounted for the error on the determination of $k$ by propagating the errors on the $g(z)$ function. This naturally increases the associated uncertainties on the luminosities. The correction for $g(z)$ affects the spread of the luminosities, hence the dispersion of the correlation, which is consequently larger.
To summarize, the dispersion increases due to the larger spread of the luminosities and it is minimally affected by the error propagation due to $g(z)$. In other words, the dispersion yielded by the function $g(z)$ is larger than the contribution given by the additional errors due to $g(z)$. Larger errors on the variables may reduce the dispersion, but in this case not sufficiently enough to balance the increase of the dispersion due to the function $g(z)$.

We also would like to point out that in a very recent paper, \cite{2021ApJ...914L..40D}, it has been shown that this method is reliable regardless of the choice of the limiting values for several sample sizes for Short GRBs (samples of 56, 34 and 32 GRBs). Thus, the discussion of \cite{2021MNRAS.504.4192B} on the EP method and its applicability are not a concern given the approach and the reliability of the results in \cite{2021ApJ...914L..40D}. 

\subsection{Impact of cosmology on the $g(z)$ function.}\label{impact on cosmology of the g(z)}

\begin{figure}
    \centering
    \includegraphics[width=0.49\textwidth]{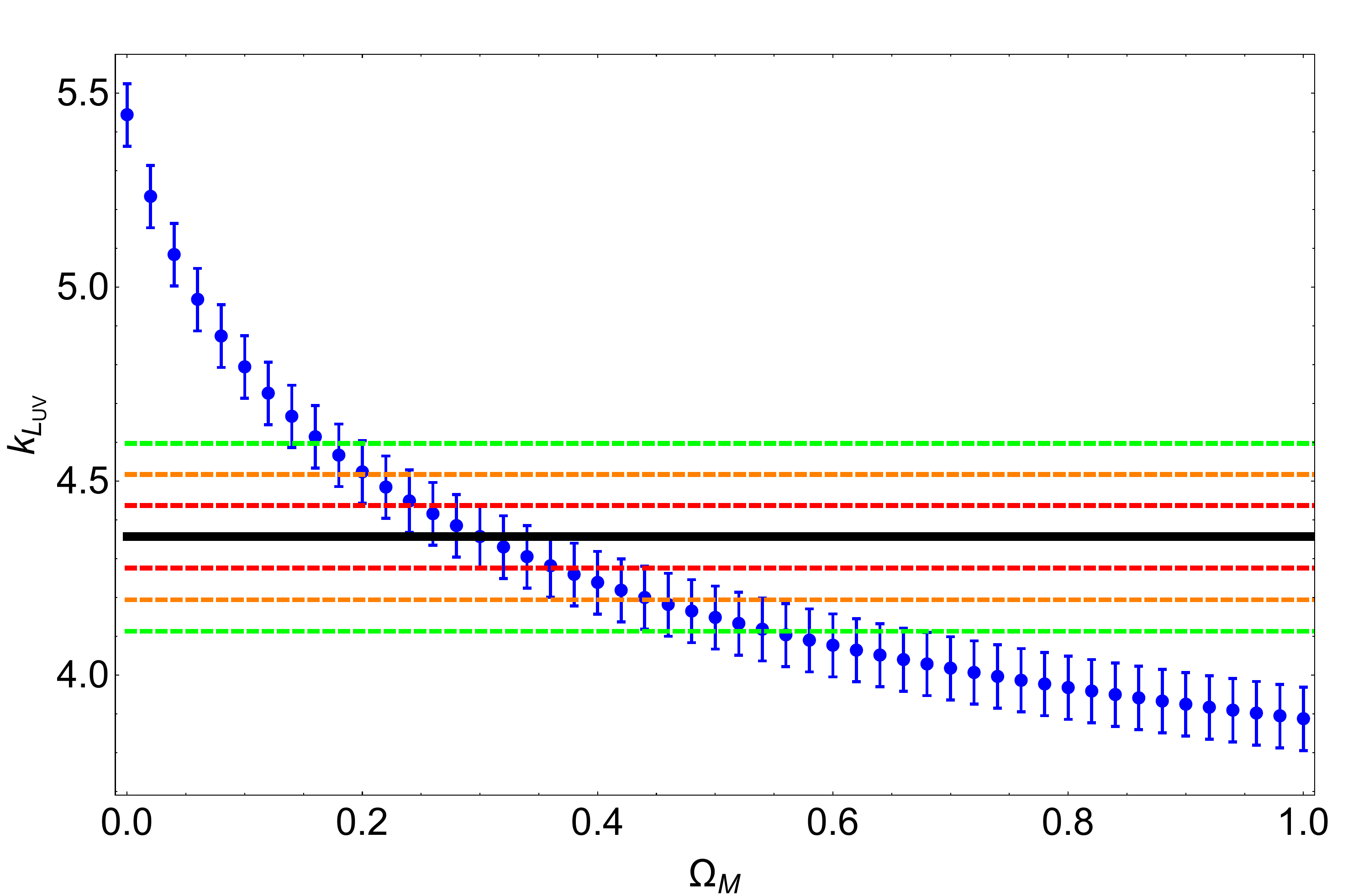}
    \includegraphics[width=0.49\textwidth]{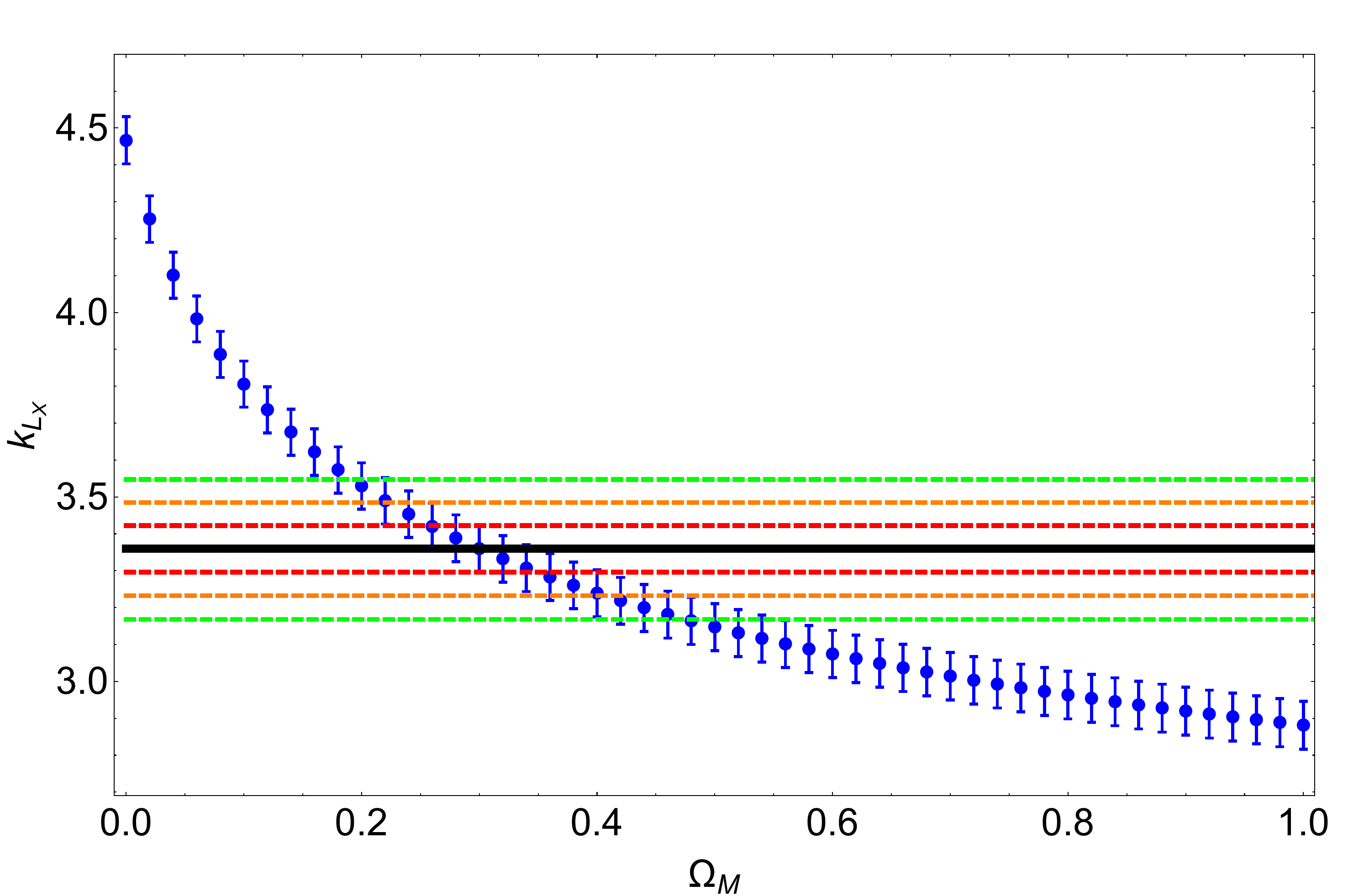}
    \caption{The dependence on the evolutionary parameter k for the $L_{UV}$ (left panel) and $L_X$ (right panel) luminosities on $\Omega_M$. The 1, 2, and 3 $\sigma$ ranges of the values of k are shown in red, orange and green, respectively. The black line indicated the value of k for $\Omega_M=0.3$ indicated in the text.}
    \label{fig:evoOMUV}
\end{figure}

In order to compute the evolutionary parameter, $k$, for the luminosities one has to assume initial fiducial values of cosmological parameters. This could possibly lead to circularity problem in cosmological measurements. We investigate the relation between the evolutionary parameter $k$ and cosmology by repeating the evaluation of the $k$ parameter following the same procedure over a set of 50 $\Omega_{M}$ values ranging from 0 up to 1. Results of this computation are shown in Fig. \ref{fig:evoOMUV}. We note here that there is no change in the value of $k$ when $H_{0}$ is varied. This happens because of the relation between $H_{0}$, the luminosity and redshift. Hubble constant is responsible only for an overall scaling of the distribution of the luminosities according to Equation (2). This does not change the number of associated sets for each redshift since both the luminosities and the limiting luminosities are scaled in the same way through the distance luminosity. Thus, there is no impact of $H_0$ on $k$. The behaviour of the $k$ parameter as a function of $\Omega_{M}$ is not negligible in a wide range of investigated values, but its values remain compatible within 1 $\sigma$ for $\Omega_{M}=0.3$ for the sets of $\Omega_{M} \in (0.20,0.45)$ and $\Omega_{M} \in (0.22,0.41)$ for cases of $k_{L_{UV}}$ and $k_{L_{X}}$ respectively. The 1, 2 and 3 $\sigma$ ranges are shown in red, orange and green. The black line indicates the value of k for $\Omega_M=0.3$ which is our reference value given that we correct the Risaliti-Lusso with a $g(z)$ based on $\Omega_M=0.3$. These ranges of values exceed the values of the most up-to-date cosmological measurements of $\Omega_{M}$ with SNe Ia ($\Omega_{M}=0.298 \pm 0.022$ \citealt{scolnic2018}) within error bars. Thus, we do not expect this effect to have a significant impact on cosmological constraints. We note that this study is important for any probe and it will be included in future analysis when QSOs are applied as cosmological probes. This method is very general and it can be also applied for any astrophysical sources observed at cosmological redshifts.

\section{The Intrinsic $\mathrm{log}L_{X}$ - $\mathrm{log}L_{UV}$ correlation}
\label{correlation}

Having overcome the impact of selection biases and redshift evolution, we can now test whether the UV-X Risaliti-Lusso relation still holds between the de-evolved luminosities we computed.
We compute the fitting parameters through a Bayesian technique, the D'Agostini method \citep{2005physics..11182D} to check if the new parameters devolved, of the normalization, $\beta^{'}$, and the slope, $\gamma^{'}$ are consistent within 1 $\sigma$ with the parameters not evolved, $\beta$ and $\gamma$. We additionally use the Python package emcee \citep{2013PASP..125..306F} to further verify our results. These method have the advantage of accounting for both errorbars on x and y axes and also an intrinsic dispersion $\delta$. The likelihood used in the D'Agostini procedure is the following: 
\begin{equation}
\displaystyle
\frac{1}{2}\sum_{i}\mathrm{ln}[\delta'^2+\gamma'^2\,\sigma^2(\mathrm{log}L_{UV,i})+\sigma^2(\mathrm{log}L_{X,i})]
			+\frac{1}{2}\sum_{i}[\gamma'\, \mathrm{log}(L_{UV,i}) + \beta'-\mathrm{log}(L_X,i)]^2/[\delta'^2+\gamma'^2\,\sigma^2(\mathrm{log}L_{UV,i})+\sigma^2(\mathrm{log}L_{X,i})]
\end{equation} 
where $\sigma(\mathrm{log}L_{UV,i})$ and $\sigma(\mathrm{log}L_{X,i})$ are the uncertainties on the UV and X-ray fluxes, respectively.

\begin{figure*}[h!]
\plotone{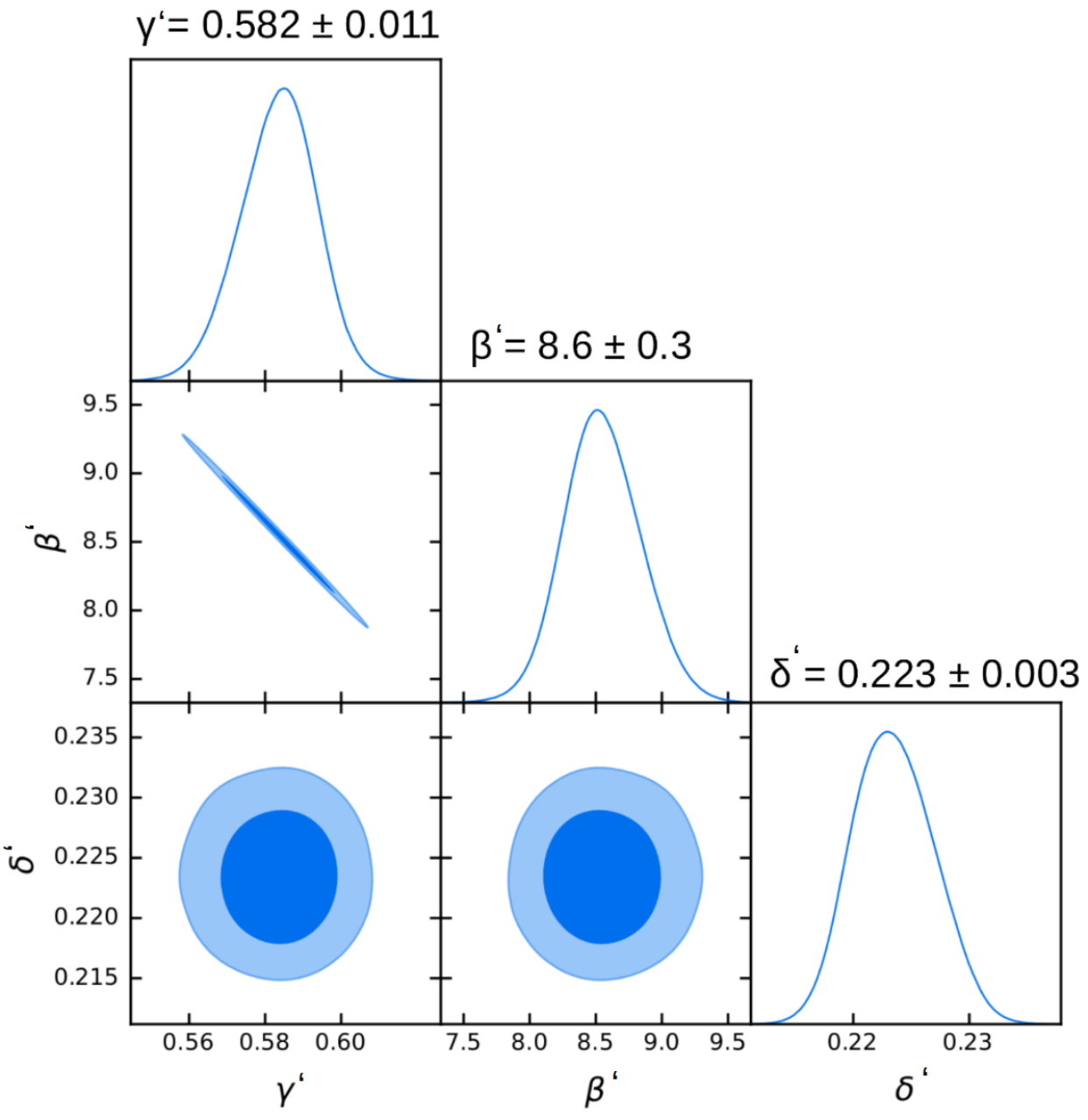}
    \caption{Results from D'Agostini fit method assuming a linear model $\mathrm{log} L'_{X} = \gamma' \, \mathrm{log} L'_{UV} + \beta'$ for the de-evolved luminosities with intrinsic dispersion $\delta'$.} 
    \label{fit}
\end{figure*}

The D'Agostini methods or similar ones are the most suitable as in our case we expect an intrinsic scatter in the UV-X relation and the error bars on both variables are not negligible. These two fitting techniques give completely consistent results within 1 $\sigma$. Assuming a linear model of the form $\mathrm{log} L'_{X} = \gamma^{'} \times, \mathrm{log} L'_{UV} + \beta^{'}$ with intrinsic dispersion $\delta$, the resulting best-fit values for the free parameters and their associated 1 $\sigma$ uncertainties from the D'Agostini fit method are: $\gamma^{'} = 0.582 \pm 0.011$, $\beta^{'} = 8.6 \pm 0.3$ and $\delta = 0.223 \pm 0.003$. We show the corner plot corresponding to these values in Fig.\ref{fit}, where the covariance between $\gamma^{'}$ and $\beta^{'}$ is just a mere effect of the fact that we perform the fit without normalizing the variables.

Given our results, we have proven that the correlation is intrinsic to the physics of QSOs and not an artifact of selection biases and/or redshift evolution and that it can be used to turn QSOs into reliable cosmological probes.
Remarkably, evolutionary parameters derived from the simple power-law or the more complex function lead to the same results for the intrinsic slope of the correlation. While  
the simple power-law $g(z) = (1+z)^k$ yields significantly different values for $k$  in  the UV and X-ray analyses, with a discrepancy of 4.4 $\sigma$ and 5.3 $\sigma$ respectively, it  leads to values for $\gamma^{'}$, $\beta^{'}$, and $\delta^{'}$ parameters consistent within 1 $\sigma$ with the values obtained from our other evolution function, as detailed in Table \ref{tab:g(z) test}. Thus, we have shown that the relation is reliable against the specific choice of $g(z)$ in the EP method.
Therefore, any approach that involves the use of this correlation to derive cosmological parameters should take into account the evolutionary function for the luminosities, otherwise we could possibly see a trend of varying $\beta$ and $\gamma$ due to the fact that the evolution has not been removed.

To test the reliability of our results we check if our fitting model assume the scatter about the line to be Gaussian. We perform this test with both Anderson-Darling and Shapiro-Wilk normality tests on the whole QSO sample, but we do not recover a normal distribution. On the other hand, we recover it if we apply a $3 \,\sigma$ clipping on the sample while fitting the linear relation. 
This procedure removes iteratively the outliers from the fitting at a chosen value of the $\sigma$ from the fitting itself. In our case, we choose 3 $\sigma$.
Specifically, the sigma-clipping procedure removes only $1.2\%$ of sources and remarkably does not change the best-fit values for the slope and the normalization, while it removes possible outliers. This procedure, which is an iterative method, has been previously reliably applied to this relation by \citet{2021A&A...649A..65B} and is commonly used when using QSOs for cosmological applications. Applying the two normality tests on this new cut sample, we recover a Gaussian distribution as we get $p_{value}=0.18$ from the Shapiro test and that the null hypothesis that the sample comes from a normal distribution cannot be rejected at more than $15\%$ significance level from the Anderson test.
\begin{deluxetable*}{ccclDlc}
\tablenum{1}
\tablecaption{Best-fit values and 1 $\sigma$ uncertainties of $\gamma'$, $\beta'$ and $\delta'$ assuming two different functional form for $g(z)$\label{tab:g(z) test}}
\tablewidth{0pt}
\tablehead{
\colhead{$g(z)$} & \colhead{$\gamma'$} & \colhead{$\beta$'} & \colhead{$\delta'$}\\
}
\startdata$(Z^k \,\mathrm{x}\, 3.7^k)/(Z^k+3.7^k)$ & $0.582 \pm 0.011$ & $8.6 \pm 0.3$ & $0.223 \pm 0.003$\\
$Z^k$ & $0.578 \pm 0.010$ & $8.8 \pm 0.3$ & $0.223 \pm 0.004$\\
\enddata
\tablecomments{For sake of clarity $Z=1+z$, as in the main text.}
\end{deluxetable*}

The best-fit values of $\gamma^{'}$ and $\delta^{'}$ completely agree with the most recent ones reported in \citet{2020A&A...642A.150L}, that are obtained with a different approach in which de-evolved observables are not taken into account. 
Indeed, in that work, the $F_X-F_{UV}$ relation is fitted in redshift bins so narrow that the spread in luminosity distance can be neglected compared to the intrinsic dispersion of the relation and, thus, fluxes can be considered as proxies for luminosities. This way, \citet{2020A&A...642A.150L} establish the non-evolution of the slope $\gamma$ with the redshift and find $\gamma = 0.586 \pm 0.061$ and $\delta = 0.21 \pm 0.06$ with a simple forward fitting method. We stress that the method adopted here is completely non-parametric.
To summarize, we show that $\gamma^{'}$, since it has no statistically meaningful change compared to $\gamma$, undergo no evolution consistently to what it is shown in Fig. 8 of \citet{2020A&A...642A.150L}.

In the \cite{2020A&A...642A.150L} analysis, the intercept fitted is not $\beta$ itself, but a combination of $\beta$, $\gamma$ and the luminosity distance, so we cannot make an immediate comparison. Nevertheless, our value of $\beta^{'}$ after the correction for evolution is exactly the one expected from the literature (i.e. $\beta \sim 8.5$) and the latest works on QSOs as standardizable candles \citep[see e.g.][]{rl19,2020A&A...642A.150L,2021arXiv210903252B}. These results establish that the Risaliti-Lusso relation is intrinsic to the the physical processes in QSOs and is not a consequence of possible selection biases and redshift evolution, as it does not change once we have removed them.
Other works have investigated several source of selection biases due to the change of the viewing angle \citep{2021ApJ...909...58P}, but in successive paper there was not found any such trend in the data \citep{2022ApJ...925..215P}. This supports the argument for no evident or not yet investigated further bias effects.

\section{Summary \& Conclusions}
\label{conclusion}

In this work, we tested the reliability of the Risaliti-Lusso relation used to standardize QSOs as cosmological candles against selection biases and redshift evolution. With this aim, we applied the \citet{1992ApJ...399..345E}, statistical method, which is specifically designed to overcome these effects, to the sample of 2421 QSOs described in \citet{2020A&A...642A.150L}. More precisely, we identified the flux limit to be applied to both the measured UV and X-ray  fluxes on the basis that it preserved at least 90\% of the original population and ensures a good resemblance to the overall distribution. In particular, we chose $F_{lim} = 4.5\times 10^{-29} \,\mathrm{erg \, s^{-1} \, cm^{-2} \, Hz^{-1}}$ for the UV and $F_{lim} = 6\times 10^{-33}\, \mathrm{erg \, s^{-1} \, cm^{-2} \, Hz^{-1}}$ for the X-rays. Using data sets trimmed to these values of the fluxes, we then found the redshift evolution of UV and X-ray luminosities under the assumption of a functional form $(Z^k \, z^k_{crit})/(Z^k+z^k_{crit})$, with $Z=1+z$ and $z_{crit}=3.7$ \citep{2013ApJ...764...43S,2016ApJ...831...60S,2019ApJ...877...63S}, using an adaptation of the Kendall $\tau$ statistic, as described in Sect.\ref{method}. This method provides us with $k= 4.36 \pm 0.08$ and $k=3.36 \pm 0.07$ for UV and X-ray evolution, respectively. These evaluations of the evolutionary coefficients allow us to compute the de-evolved luminosities $L'$ for the whole original sample, which we fit with two different techniques assuming the linear model $\mathrm{log} L'_{X} = \gamma' \, \mathrm{log} L'_{UV} + \beta'$. The latter is exactly the same form as the Risaliti-Lusso relation, but using de-evolved luminosities. We obtained completely consistent results from all the fitting methods applied; specifically from the D'Agostini fit $\gamma^{'} = 0.582 \pm 0.011$, $\beta^{'} = 8.6 \pm 0.3$ and $\delta^{'} = 0.223 \pm 0.003$, where $\delta^{'}$ is the intrinsic dispersion of the relation and we quote 1 $\sigma$ uncertainties. 

These results are independent from changes in the specific choice of the functional form assumed for the evolution and from the initial choices of the limiting fluxes for more than one order of magnitude, as we have demonstrated by testing also the case of a simple power-law (see Table \ref{tab:g(z) test}) and for different flux limits.

In summary, the values of the slope and the normalization obtained completely agree with results from the literature 
on QSOs as standardizable candles \citep[see e.g.][]{rl19,2020A&A...642A.150L,2021arXiv210903252B}, which makes use of completely different methodology. This shows that the Risaliti-Lusso relation persists once selection biases and redshift evolution are removed, and as a result, it is completely intrinsic to the physics of QSOs. In conclusion, 
the outcome of this investigation paves the way to new routes on the possibility to standardize quasars through the Risaliti-Lusso relation as cosmological candles, thereby extending the Hubble diagram up to $z=7.54$.

\section{acknowledgments}
MGD acknowledges the Division of Science and NAOJ for the support. 
E\'OC was supported by the National Research Foundation of Korea grant funded by the Korea government (MSIT) (NRF-2020R1A2C1102899). DS is partially supported by the US National Science Foundation, under Grant No. PHY-2014021. MMShJ would like to acknowledge SarAmadan grant No.
ISEF/M/400121.

%
%
%

\appendix

\section{Details on the EP method}
\label{appendix}
To ease the reader in navigating through the statistical EP method we here include a more mathematical background on this procedure.
Truncated data plays a crucial role in the statistical analysis of astronomical observations. Indeed, the original example quoted in the EP method concerns a set of measurements on QSOs.
The EP method allows to overcome the problem of truncation effects with a two-dimensional extension of the unbinned Lynden-Bell's C- method \citep{1971MNRAS.155...95L}, rediscovered by statisticians Woodroofe \citep{notfound1}, Wang \citep{notfound2}, Petrosian \citep{1992scma.conf..173P}, and Efstathiou \citep{1988MNRAS.232..431E}. 

The Lynden-Bell-Woodroofe-Wang method is established by a theorem that this is the unique non-parametric maximum likelihood estimator of randomly truncated univariate data, analogous (and mathematically similar) to the famous Kaplan-Meier estimator \citep{notfound3} for randomly censored data. This means that there is no better estimator available if the assumptions hold. In addition to the Efron-Petrosian studies, it is the foundation of other methods such as a two-sample test, correlation, linear regression, and Bayesian analysis. Details appear in \citet{notfound4}.

Although we have preferred to develop our own code in Mathematica, we here acknowledge that there are public codes for the LBW estimator, its confidence bands, Efron-Petrosian, and related procedures given in CRAN packages 'DTDA' 'double.truncation' in the public domain R statistical software environment (A cookbook is provided by \citealt{notfound4}).

A simple example of this method appears in the textbook \citet{notfound5}.
In this case the statistic $V / V_{max}$ method, first introduced by \citet{1968ApJ...151..393S}, underestimates the LF in the fainter bins. The $V / V_{max}$ is a measure of the uniformity of the space distribution of sources (initially applied for radio sources later has been applied to X-ray sources too). This method is based on the ratio of the volume ($V$) enclosed at the redshift of an object divided by the volume ($V_{max}$) that would be enclosed at the maximum redshift at which the object would be detectable. The Luminosity Function measured with the $V/V_{max}$ method performs less well (an example is shown in \citealt{notfound5}) compared to the Lynden-Bell estimator. For a more recent discussion of one dimensional luminosity function methods see \citet{2013Ap&SS.345..305Y}.


\bibliography{bibliografia}{}
\bibliographystyle{aasjournal}



\end{document}